\documentclass[prl,aps,onecolumn,notitlepage,prb,preprintnumbers,amsmath,amssymb,aps,floatfix]{revtex4}

\usepackage{amsmath,amssymb,mathtools,epsfig,epstopdf,mathpazo,textcomp,graphicx,chemarrow,soul}
\usepackage[dvipsnames]{xcolor}
\usepackage[toc,page]{appendix}

\usepackage{txfonts}

\newcommand{\be}{\begin{eqnarray}}
\newcommand{\ee}{\end{eqnarray}}

\begin{document}

\title{Bimodality in gene expression without feedback:\\ From Gaussian white noise to log-normal coloured noise}

\author{Gerardo Aquino$^1$ and Andrea Rocco$^{1,2}$}
\email{a.rocco@surrey.ac.uk}

\affiliation{$^1$Department of Microbial Sciences, Faculty of Health and Medical Sciences, University of Surrey, GU2 7XH Guildford, United Kingdom\\
  $^2$Department of Physics, Faculty of Engineering and Physical Sciences, University of Surrey, GU2 7XH Guildford, United Kingdom}

\begin{abstract}
  Extrinsic noise-induced transitions to bimodal dynamics have been largely investigated in a variety of chemical, physical, and biological systems. In the standard approach in physical and chemical systems, the key properties that make these systems mathematically tractable are that the noise appears linearly in the dynamical equations, and it is assumed Gaussian and white. In biology, the Gaussian approximation has been successful in specific systems, but the relevant noise being usually non-Gaussian, non-white, and nonlinear poses serious limitations to its general applicability. Here we revisit the fundamental features of linear Gaussian noise, pinpoint its limitations, and review recent new approaches based on nonlinear bounded noises, which highlight novel mechanisms to account for transitions to bimodal behaviour. We do this by considering a simple but fundamental gene expression model, the repressed gene, which is characterized by linear and nonlinear dependencies on external parameters. We then review a general methodology introduced recently, so-called nonlinear noise filtering, which allows the investigation of linear, nonlinear, Gaussian and non-Gaussian noises. We also present a derivation of it, which highlights its dynamical origin. Testing the methodology on the repressed gene confirms that the emergence of noise-induced transitions appears to be strongly dependent on the type of noise adopted, and on the degree of nonlinearity present in the system.
\end{abstract}


\maketitle

\section{1. Introduction}
Multistability, the simultaneous existence of two or more stable steady states, is a feature dramatically relevant for the functioning of living systems \cite{Pomerening08}. Deterministic multistability is realized in gene regulatory networks by the inclusion of specific topological features. Feedback loops for instance are known determinant of multistable behaviours \cite{Leite10,Balazsi11,Wang19} and underlie a number of processes involved in cellular decision-making, ranging from the LAC operon \cite{Santillan07}, to quorum sensing \cite{Williams08,Melke10}, to cell differentiation \cite{Ferrell12}.

Exploration of multiple deterministic steady states is realized through stochastic fluctuations at the molecular level, or noise. Noise is ubiquitous in molecular biology, and occurs in many aspects of gene regulation and other cellular processes in bacteria \cite{Elowitz02}, yeast \cite{Gasch17}, and mammalian cells \cite{Ochiai19}. By allowing transitions among alternative steady states, it accounts for much if not all of the variability that we see in biological systems \cite{Raser05,Eldar10,Tsimring14}. 

In general, noise is classified in the two broad categories of intrinsic and extrinsic noise. Intrinsic noise originates in the natural randomness of molecular events, and affects directly the dynamical variables of the system, such as for instance protein or mRNA concentrations of a gene regulatory network. Its modelling is based on a wealth of results, which aim to solve the Chemical Master Equation \cite{vanKampen07}, analytically for instance in terms of the Linear Noise Approximation \cite{Elf03}, or by direct stochastic simulations, using the celebrated Gillespie algorithms  \cite{Gillespie77,Gillespie01}.

Extrinsic noise in contrast represents the effect of a fluctuating environment on the system. It translates mathematically into stochastic fluctuations acting on the parameters that enter the system, such as pH levels, temperature, or rate and degradation constants. Extrinsic noise too has been largely investigated in biology in the last two decades \cite{Swain02,Dattani16,Bressloff19}.

Interestingly, extrinsic noise has been shown to produce highly non-trivial effects, responsible for the emergence of multistability or oscillatory behaviours. New mechanisms, purely stochastic, have now been identified as determinants of multistable behaviour, in addition and possibly in alternative to the selection of deterministic tolopogical features. These effects are referred to as 'purely' noise-induced transitions, where the term 'purely' is used to emphasize the difference with respect to transitions driven by noise among deterministically pre-exisiting steady states, and is often omitted in the literature once the context is clear. We will also omit it in the rest of this paper. 

Two fundamentally different mechanisms are known to cause noise-induced transitions. The first mechanism refers to the fact that in presence of fluctuating environments the stochastic differential equation (SDE) describing the system contains multiplicative noise terms. If Gaussian white noise is chosen to describe the environmental fluctuations, the presence of these terms makes the problem mathematically ill-defined, and the solution of the corresponding SDE is not unique. To make the problem mathematically well-defined and tractable, the SDE requires a specific discretization prescription that defines its stochastic integral \cite{vanKampen07,Horsthemke84,Gardiner85}. The It\^{o} and the Stratonovich prescriptions have been put forward as sensible choices to compute a unique solution of the SDE describing the system, but which prescription is the 'right' one has been debated for a long time. It is now widely recognized that the requirement that the noise be charaterized by a small but finite correlation time implies that the Stratonovich prescription needs to be adopted in the corresponding white noise limit. We call this type of noise 'effectively white' or 'physical' noise, in that its frequency spectrum is characterized by a high frequency cut-off, as indeed happens for physical noises. The adoption of the Stratonovich prescription then leads to the emergence of a spurious drift, the so-called Stratonovich drift, that modifies the deterministic dynamics of the system, and gives rise to non-trivial properties and noise-induced transitions.

The emergence of noise-induced transitions relying on the Stratonovich drift has been the subject of intense theoretical and experimental research in a number of different physical and chemical systems, including homogeneous \cite{Horsthemke84} and spatially extended systems \cite{Sancho99}, but also in ecological modelling \cite{Zeng15}, and in socio-economic systems \cite{Jungeilges17}.

In biology, several hypotheses that rely on Gaussian white noise and on the Stratonovich drift have been put forward to explain the emergence of bimodal behaviours in sytems deterministically monostable. For instance, Samoilov et al. \cite{Samoilov05,Samoilov06} have shown that futile enzymatic cycles, deterministically monostable, can transition to a bimodal regime when noise affects the enzymes’ activities in an appropriate manner. Similar considerations have led to the formulation of the concept of stochastic control in metabolic pathways \cite{Rocco09}. In development, Pujadas and Feinberg \cite{Pujadas12} have speculated that noise participates to the developmental program, by contributing to the process of cellular decision making through modifications of the morphology of the epigenetic landscape by noise-induced transitions. In bacteria, quorum sensing \cite{Weber13a} and simple repressive circuits \cite{Weber13b} have been investigated stochastically, and have been proven to show non-trivial emerging properties related to noise-induced transitions. In all these studies, Gaussian white noise has been assumed.

However, a second mechanism accounting for noise-induced transitions has been recently discovered, formalized theoretically, and verified experimentally. Ochab-Marcinek et al. \cite{Ochab-Marcinek10,Ochab-Marcinek17} have proposed a neat geometric construction based on nonlinear noise processing. In its simplest version, this mechanism relies on the existence of a formal relation between two stochastic variables that mimics a nonlinear input-output response between them. Simple conservation of probabilities then establishes the output probability distribution in terms of the input probability distribution. Because of the assumed nonlinearity of the response function, it may happen that the output distribution becomes bimodal, even though the input distribution was unimodal.

This second mechanism can be formulated for any input probability distribution, not necessarily Gaussian, and for a variety of biochemical systems. For instance, Birtwistle et al. \cite{Birtwistle12} have shown that bimodality of extracellular signal-regulated kinase (ERK) emerges as a response to epidermal growth factor (EGF) in presence of a fluctuating activation threshold that can be analyzed in terms of the nonlinear noise filtering methodology \cite{Ochab-Marcinek10}, as highlighted by Kim and Sauro \cite{Kim12}. Similarly, Dobrzy{\'n}ski et al. \cite{Dobrzynski14} have studied a stochastic model of the hypoxia-inducible factor (HIF) pathway, and found that it shows emergence of bimodal dynamics of HIF depending on response threshold variability to dimethyloxalylglycine, which mimics oxygen limited conditions. What is noticeable in these studies is that they provide experimental validation not only of the occurrence of bimodal dynamics, but also that these emerge as purely induced by noise. In a different context, the differential and biphasic tolerance of clonal bacterial cells to drugs, so-called bacterial persistence \cite{Balaban04,Lou08,Hingley-Wilson20}, has also been described in terms of nonlinear noise filtering \cite{Rocco13a,Rocco13b}.

It is interesting to note that the geometric construction presented in \cite{Ochab-Marcinek10} can be derived in dynamical terms  by extending to nonlinear noise a well established formalism, first developed in \cite{Arnold78} and applied later in a variety of chemical and physical systems \cite{Reichl82}. The formalism is based on relaxing the assumption of white noise, and on focusing instead on fluctuations characterized by a correlation time much longer than all other relaxational timescales in the system. We present here this derivation, which offers further insights into the nature of the related noise-induced transitions, and allows for a direct comparison with the Stratonovich approach.

Despite its success over the years, and independently of the mechanism responsible for the occurence of noise-induced transitions, the Gaussian white noise assumption is not free of criticism in gene regulatory networks. First, the Gaussian approximation is of limited applicability for fluctuations affecting parameters which are strictly positive. Second, the white noise assumption is also problematic in GRNs. From the biological perspective, the white character of extrinsic noise is largely questionable for gene regulatory networks, where the typical correlation times of extrinsic noise can extend up to and well above typical cell cycle times \cite{Rosenfeld05,Kaufmann07}. Third, regulatory dynamics described mathematically in terms of Hill functions include nonlinear dependencies on some of the parameters. This makes the mathematical treatement of the corresponding stochastic dynamics for white noise very hard, if not impossible \cite{Horsthemke84}.

These problems have been recognized in some recent literature, and relevant progress has been made in characterizing the response of different systems under more biologically grounded  noise. Non-white noise, so-called coloured noise, characterized by a finite correlation time, can be addressed by the powerful approach of the Unified Colored Noise Approximation (UCNA) \cite{Jung87}, which captures very well the dynamics of the system under the Gaussian approximation. Recently, Holehouse et al. \cite{Holehouse20} have applied UCNA to a genetic feedback loop, and a thorough analysis of the effect of Gaussian coloured noise has identified noise-induced transitions from uni- to bimodal behavior and vice versa. We review the UCNA approach in the next section, as a first step in addressing the effect of finite noise correlations under the Gaussian approximation.

In order to relax the Gaussian approximation itself, so-called bounded noise has been introduced recently. Bounded noise is generally obtained by starting from a a Gaussian process, and imposing hard bounds on it, so as to constrain it in a bounded domain. Examples are the Sine-Wiener process, the Tsallis noise, or the Borland noise \cite{Borland98,Bobryk05,deFranciscis14}. These different types of noise have been studied in proof of concepts examples, such as the genetic model \cite{Wio04}, and in applied set-ups \cite{dOnofrio10a,dOnofrio10b}.  Gamma-distributed noise and log-normal noise are also of marked interest in that they are particularly successful in fitting single cell experiments \cite{Cai06,Bengtsson05}, and can be derived under very broad assumptions from the statistical properties of gene expression data \cite{Ham20}. They preserve by construction the positivity of the relevant variable, as  their domain of definition is semi-bounded, extending from zero to infinity. All these noises have been found to produce noise-induced transitions, and in general a strong divergence from the expected Gaussian behaviour.  

In this paper we review the two main mechanisms by which noise-induced transitions can emerge. First we present the standard derivation of the Stratonovich drift for Gaussian white noise and the UCNA approach for small but finite noise correlation time. Then we introduce the noise filtering approach, and show how this method can be derived dynamically by considering slow fluctuations. We consider both Gaussian and log-normal noise, and review the implications of both approaches for a simple but fundamental system, the repressed gene, analysing both linear and nonlinear dependencies. We finally discuss the main biological implications of these results, and highlight a series of open problems and challenging questions.

\section{2. The Stratonovich interpretation for Gaussian white noise and the UCNA approach}
Let us consider the following dynamical system:
\be
\frac{dx}{dt} = f(x,\lambda), \label{ODEp}
\ee
where $\lambda$ is a control parameter, related to environmental fluctuations, and which appears linearly in the dynamics specified by the function $f$. Let us assume that $R$ is affected by fluctuations
\be
\lambda \rightarrow \lambda(t) = \bar{\lambda} + D^{1/2}\xi(t). \label{setRs}
\ee
Here $\bar{\lambda}$ is the average of $\lambda$, $\xi(t)$ is Gaussian, zero-average, white noise, with intensity $D$ and correlator given by
\be
\langle \xi(t)\xi(t')\rangle = 2 \delta(t-t'), \label{corr}
\ee

Because of the linear dependency of $f$ on $\lambda$, we can decompose the fluctuations acting in (\ref{ODEp}) into two terms, and write:
\be
\frac{dx}{dt} = f(x,\bar{\lambda}) + D^{1/2} g(x) \xi(t), \label{SDEgw}
\ee
where $g(x)=\partial f(x,\lambda)/\partial \lambda$.

Eq. (\ref{SDEgw}) is a multiplicative noise equation, and it is ill-defined as it stands. Its definition relies on the time discretization adopted to evaluate its stochastic integral, and it only acquires meaning if supplemented with an adequate discretization prescription \cite{Gardiner85}. Historically two prescriptions have been formulated and debated, the It\^{o} prescription and the Stratonovich prescription, which are implemented by considering the two following versions of Eq. (\ref{SDEgw}):
\be
\frac{dx}{dt} = f(x) + D (2-\alpha) g(x) g'(x) + D^{1/2} g(x) \xi(t), \label{SDEI}
\ee
Setting $\alpha=2$ corresponds to assuming the It\^{o} prescription, while setting $\alpha=1$ corresponds to assuming the Stratonovich prescription. The term $D g(x) g'(x)$ in (\ref{SDEI}) and (\ref{fp}) is the so-called 'Stratonovich drift', and is responsible for the emergence of non-trivial dynamics, as we shall discuss. The associated Fokker-Planck equation reads
\be
\frac{\partial p(x,t)}{\partial t} = -\frac{\partial}{\partial x}\big[\big(f(x) + D (2-\alpha) g(x)g'(x)\big)p(x,t)\big]
+ D \frac{\partial^2}{\partial x^2}\big[g^2(x)p(x,t)\big]. \label{fp}
\ee 
Eq. (\ref{fp}) can be solved for the stationary probability distribution \cite{Gardiner85}:
\be
p(x) = \frac{\cal{N}}{g^{\alpha}(x)} \exp\left[\frac{1}{D}\int^x \frac{f(y)}{g^2(y)}dy\right] \label{stp}
\ee
where $\cal{N}$ is a normalization factor. The modes of the stationary probability distribution correspond to the stochastic extension of the steady states of the system, and are given by
\be
f(x) - D \alpha g(x)g'(x)=0. \label{modes}
\ee
Eq. (\ref{modes}) clearly shows that depending on the interplay between the functions $f(x)$ and $g(x)$, the modes can be dramatically different from the deterministic steady states of the system, identified by the equation $f(x)=0$. This is in principle true for both the It\^{o} and the Stratonovich prescription, as the Stratonovich drift appears in Eq. (\ref{modes}) in both cases, albeit with a different weight. Its effect is to possibly change the number and the stability properties of the deterministic steady states, with the noise intensity $D$ acting as a bifurcation parameter, driving transitions between unimodal to bimodal or multimodal dynamics. Because these transitions are absent in the deterministic system, they are called noise-induced transitions. The noise is not only driving transitions among pre-existing deterministic steady states, but it possibly creates new ones, and drives transitions among them.\\

Whether the It\^{o} or the Stratonovich prescription is to be assumed, has been extensively debated in the past. While the two prescriptions are mathematically legitimate and equivalent, of course the implications can be different on noise-induced transitions, as the $\alpha$-dependency in Eq. (\ref{modes}) shows. For physical and chemical systems, the choice depends on the way noise is derived and on the way if affects the system itself \cite{vanKampen07}. For biological systems, similar considerations apply, with different conclusions being drawn according to the type of noise that is examined. For instance, intrinsic noise can be modelled effectively in terms of the Chemical Langevin Equation \cite{Gillespie92,Gillespie00}, which needs to be interpreted in the sense of It\^{o}, and can be shown to lead to noise induced transitions in small gene regulatory circuits \cite{Weber13b}. It should be emphasized that the appropriate definition of Langevin dynamics has to rely on the definition of the corresponding master equation, as is done in \cite{Gillespie92,Gillespie00}, whereas arbitrary insertion of noise to mimic intrinsic fluctuations leads to unphysical and inconsistent results \cite{vanKampen07}. In contrast, when extrinsic noise is considered, the assumption that white noise is the limit of Gaussian noise with small but finite correlation time, and therefore 'physical', requires the adoption of the Stratonovich prescription \cite{Wong65}.

In order to explore further the behaviour of the system when driven by Gaussian noise with small but finite correlation time, let us consider the following dynamical system:
\be
&&\frac{dx}{dt} = \frac{1}{\tau_x}f(x,\lambda) \label{redPn}\\
&&\frac{d\lambda}{dt} = - \frac{\lambda}{\tau} + \frac{\sqrt{D}}{\tau} \xi. \label{noiseOU}
\ee
In Eq. (\ref{redPn}) the relevant timescale for $x$, $\tau_x$, is now explicity written for clarity. In contrast to the white noise case, with Eq. (\ref{noiseOU}) we now assume that the parameter $R$ is described by a Ornstein-Uhlenbeck process \cite{Gardiner85}, characterized by a correlation time $\tau$, driven by the Gaussian, zero average, white noise $\xi=\xi(t)$, with correlator given by (\ref{corr}). This system extends the system defined by Eqs. (\ref{ODEp}), (\ref{setRs}) and (\ref{corr}) for finite correlation time of the noise. In fact, the correlator of the $\lambda$ process can be easily calculated, and results in
\be
\langle \lambda(t)\lambda(t^\prime)\rangle = \frac{D}{\tau} e^{-\frac{|t-t^\prime|}{\tau}} \label{OUcorr}
\ee
For $\tau \rightarrow 0$, the correlator (\ref{OUcorr}) tends to the white noise limit, namely Eq. (\ref{corr}), and the system (\ref{redPn}) and (\ref{noiseOU}) becomes formally equivalent to (\ref{ODEp}), (\ref{setRs}) and (\ref{corr}).

For $\tau$ finite, the Fokker-Planck Equation has been derived within the so-called Unified Coloured Noise Approximation (UCNA) \cite{Jung87},
\be
\frac{\partial p(x,t)}{\partial t} = -\frac{\partial}{\partial x}\left[\left(\frac{f(x)}{g(x)}g_{\rm UCNA}(x) + D g_{\rm UCNA}(x)g_{\rm UCNA}'(x)\right)p(x,t)\right]
+ D \frac{\partial^2}{\partial x^2}\big[g_{\rm UCNA}^2(x)p(x,t)\big], \label{fp2}
\ee
where 
\be
g_{\rm UCNA} = g(x)[1-\tau(f^\prime(x) - f(x) g^\prime(x)/g(x))]^{-1},
\ee
and whose solution is
\be
p(x) = \frac{\cal{N}}{|g(x)|}|1-\tau g(x)(f(x)/g(x))^\prime]| \exp\left[\frac{1}{D}\int^x \!\! \frac{f(y)[1-\tau g(y)(f(y)/g(y))^\prime]}{g^2(y)}dy\right]. \label{stpucna}
\ee
This solution is exact in both the $\tau=0$ and $\tau=\infty$ cases. An extensive discussion of this result, including its limitations, is presented in \cite{Jung87,Grigolini88}. In the present context, we notice that it is immediate to verify that for $\tau \rightarrow 0$ Eq. (\ref{fp2}) reduces to (\ref{fp}) in the Stratonovich interpretation, confirming thereby the general result \cite{Wong65}. 

This latter argument lends itself well to considering also the neighbourhood of white noise, or so-called 'virtually white' noise. As long as $\tau \ll \tau_x$, we can effectively consider the noise as white, and use the white noise Fokker-Planck (\ref{fp}) instead of (\ref{fp2}), as long as we interpret it in the Stratonovich sense. In this respect, the white noise assumption, generically valid when the correlation time of the noise is much smaller than any other timescale of the system, simplifies greatly the treatment, and allows an effective approximate description.

\section{3. The repressed gene: Gaussian fast noise on degradation} 

The effect of Gaussian extrinsic noise, both white and coloured, on simple gene circuits has been object of recent studies. Holehouse at el. \cite{Holehouse20}, for instance, investigate the effect of noise on degradation rates by using the UCNA approach in a feedback loop in the bistable regime and in concomitance with intrinsic fluctuations on protein concentrations. Previously, Shahrezaei et al. \cite{Shahrezaei08} had addressed similar issues, but in a single activated gene, under the effect of log-normal noise.

Here, we revisit these results in a simplified but fundamental model of gene expression, the repressed gene. In order to ascertain the effect of extrinsic noise only, we assume that intrinsic noise, related to protein low copy numbers, is absent, and compare the cases of white noise and coloured noise with small but finite correlation time. 

\begin{figure}[t] 
\begin{center}
  \includegraphics[height=2.6 cm, angle=0]{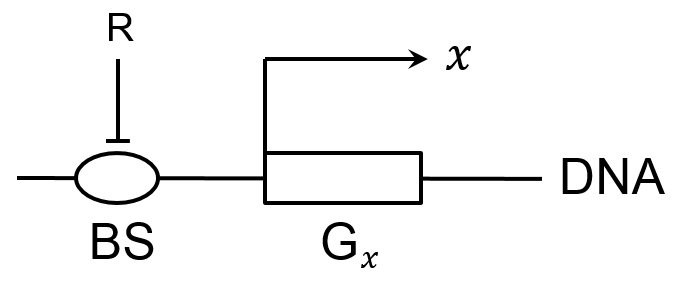}
\caption{The repressed gene. The expression level of gene $G_x$ is downregulated by the binding of the repressor $R$ to the the DNA binding site BS. We make the assumption that when active, gene $G_x$ synthesizes the protein $x$ in one single step of combined transcription and translation.}\label{repr}
\end{center}
\end{figure}

The repressed gene is a gene which is constitutively expressed at its maximal expression rate, but it is regulated by a transcription factor (TF), the repressor, that represses protein synthesis (Fig. \ref{repr}). 

We describe this system by decomposing the rate equation for the protein $x$ into a production term, described phenomenologically in terms of a Hill function, and a degradation term, straightforwardly written according to the Law of Mass Action:
\be
\frac{dx}{dt} = f(x,R,k) = \frac{g}{1 + \rho R} - k x. \label{redP1}
\ee
Here $\rho$ is the association constant of $R$ to DNA, which describes the strength of the repressor binding to its binding site, $g$ is the maximal expression rate for gene $G_x$, and $k$ is the degradation rate of the protein $x$. Due to the much shorter half-life of mRNA with respect to that of proteins \cite{Rosenfeld02}, we here assume that transcription and translation are lumped together in one single step, so that when the gene is in the active state it synthesises directly the protein $x$. Also, when the concentration of $R$ is very high, we can consider it as unaffected by the $x$ dynamics, and therefore as a constant control parameter. Hence, the dynamics specified by the function $f(x,R,k)$ is characterized by a timescale $\tau_x$, which in the specific case considered here is simply given by the protein inverse degradation rate, $\tau_x = 1/k$. 

These dynamics are supplemented with the following biologically grounded parameter values, which broadly apply to typical bacterial gene expression. We estimate the effective maximal gene expression parameter in the range $g \approx g_P g_M/k_M = 10^{-2} \div  10^{-1}$ s$^{-1}$, where $g_P$ and $g_M$ are transcriptional and translational rates respectively, and $k_M$ is the degradation rate of mRNA. This choice is compatible with a moderate to strong transcriptional and translational activity and typical mRNA degradation rates in bacteria \cite{Thattai01}. Also, we set $k = 10^{-3} \div 10^{-4}$ s$^{-1}$, which corresponds to a protein half-life ranging between minutes and a few hours \cite{Toyama13}, and a typical dissociation constant of trancription factors to DNA $\rho = 0.1 \div 10$ nM$^{-1}$ \cite{Loinger07}. We also choose $R$ in the range $R = 1 \div 100 $ nM, which together with the chosen $\rho$ gives on average a weak ($\rho R <1$) to strong ($\rho R > 1$) repression activity of $R$ on gene $G_x$. Simulations in this paper are carried out with parameters with these orders of magnitude, and specific parameter values will be indicated in the captions of the figures. 

Throughout this paper we will use the repressed gene as the preferred model to revisit the effect of different types of noises on different parameters, namely on the degradation rate $k$ and on the repressor concentration $R$. We do not consider fluctuations on $g$ as these would merely correspond to insert additive noise into the system.

Following \cite{Holehouse20,Shahrezaei08}, let us consider first white noise on the degradation rate:
\be
k \rightarrow k(t) = \bar{k} + D^{1/2} \xi(t) 
\ee
where we assume $\xi(t)$ to be Gaussian and white, with correlator given by Eq. (\ref{corr}). Since the degradation rate appears linearly in Eq. (\ref{redP1}), we can apply straigthforwardly the approach introduced in the previous section, and write the corresponding SDE as 
\be
\frac{dx}{dt} = \beta - \bar{k}x - D^{1/2}x\xi(t), \label{SDE1}
\ee
which needs to be supplemented with the Stratonovich prescription as discussed, and where we have set $\beta = g/(1+\rho R)$ for convenience.

We can directly compute the normalized solution of the Stratonovich Fokker-Planck Equation (\ref{stp}) as:
\be
p(x) = \frac{(\beta/D)^{k/D}}{\Gamma(k/D)} x^{-(1+k/D)} e^{-\frac{\beta}{D} \frac{1}{x}}\label{an1}
\ee
whose mode is located at
\be
x_M^{\rm (stoch)} = \frac{\beta}{k+D}, \label{modesspec}
\ee
as implied by Eq. (\ref{modes}), and to be compared with the deterministic steady state located at
\be
x_{ss}^{\rm (det)} = \frac{\beta}{k}.
\ee

\begin{figure}[t] 
\begin{center}
  \includegraphics[height=5.0 cm, angle=0]{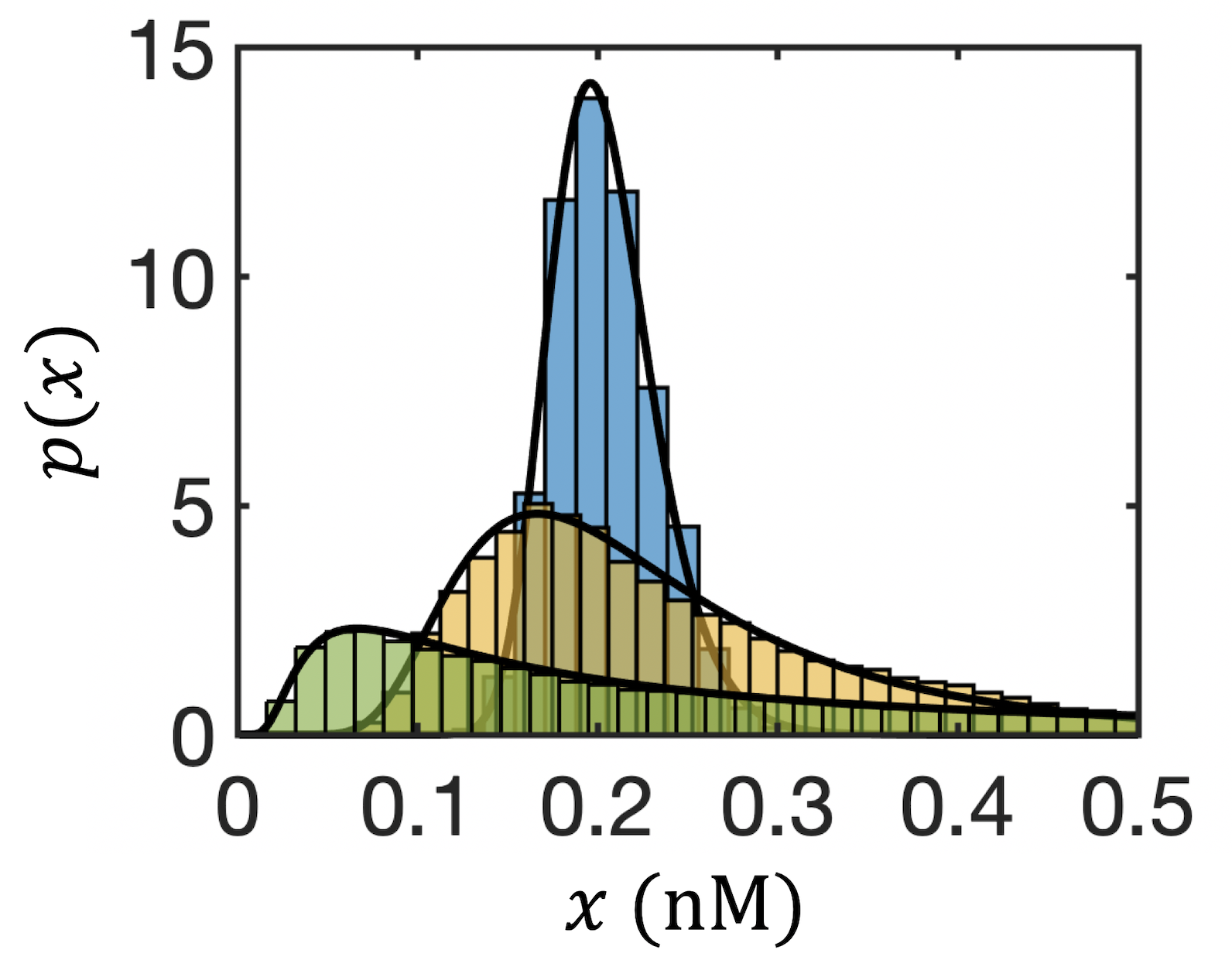}
  \caption{Probability distributions for the repressed gene with extrinsic white noise on the degradation rate $k$. The different distributions are obtained for $D =10^{-5}$ (blue), $D =10^{-4}$ (orange), $D =10^{-3}$ (green), and the full black curves represent the analytical prediction (\ref{an1}). Other parameter values are as follows: $g=0.1$ s$^{-1}$, $k=5 \cdot 10^{-4}$ s$^{-1}$, $\rho=10$ nM$^{-1}$, $R=100$ nM. With these parameters, the deterministic steady state is located at $x_{ss}^{\rm (det)} = 0.199$ nM, while the position of the modes for the stochastic system is $x_M^{\rm (stoch)} = 0.196$ nM (for $D =10^{-5}$), $x_M^{\rm (stoch)} = 0.166$ nM (for $D =10^{-4}$), and $x_M^{\rm (stoch)} = 0.066$ nM (for $D =10^{-3}$).}\label{StratFig}
\end{center}
\end{figure}

We show direct stochastic simulations of the system (\ref{SDE1}) in Fig. \ref{StratFig}. The agreement between the simulations and the analytical solution (\ref{an1}) is excellent, and it captures the predicted dynamics. Not only a broadening of the mode is observed for higher noise intensities, but a shift in the position of the mode itself takes place, which obeys the theoretical prediction (\ref{modesspec}).

\begin{figure}[t] 
\begin{center}
  \includegraphics[height=5.0 cm, angle=0]{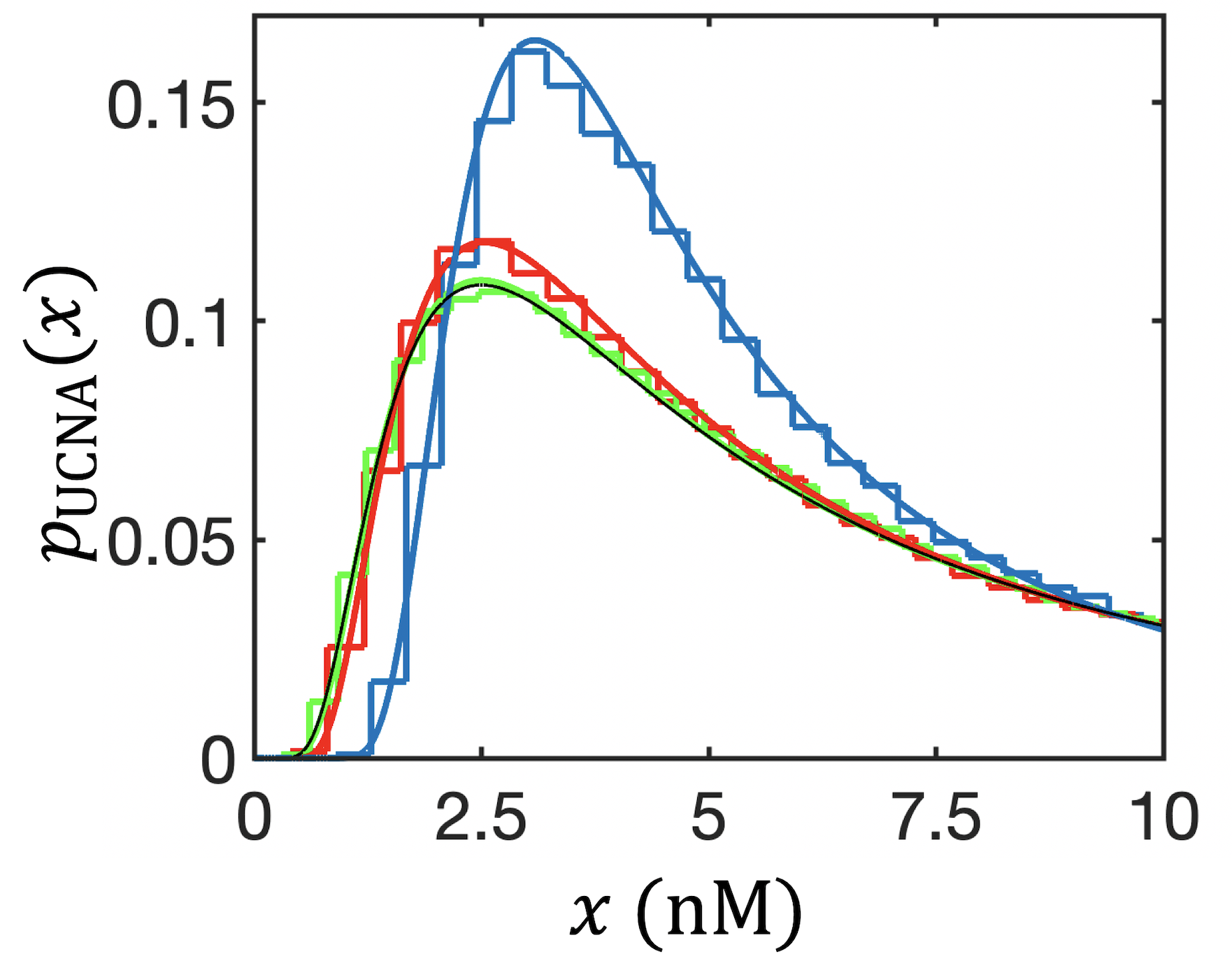} 
  \caption{UCNA probability distributions for the repressed gene with coloured extrinsic noise on the degradation rate and different $\tau$ values. Parameters are as follows: $g=0.01$ s$^{-1}$, $k=10^{-3}$ s$^{-1}$, $\rho=1$ nM$^{-1}$, $R=1$ nM $D=10^{-3}$. The thin black line corresponds to the white noise solution, while the green, red, and blue lines and histograms correspond to the UCNA solution with $\tau=10, 100, 1000$ s respectively. It is evident that the UCNA solution reproduces very well the coloured noise simulations, and that for $\tau=10$ the UCNA and the white noise solution are practically indistinguishable.}\label{UCNAFig}
\end{center}
\end{figure}

We also explore the effect of coloured noise on the system, by adopting the UCNA approach presented in the previous section, and adopted in \cite{Holehouse20,Shahrezaei08}. The UCNA solution (\ref{stpucna}) reads in this case
\be
p_{\rm UCNA}(x) = \frac{{\cal N} (x + \tau \beta)}{x^{2+k/D}} e^{-\frac{\beta}{Dx} + \frac{\tau}{D}\left(\frac{k \beta}{x} - \frac{\beta^2}{2 x^2}\right)},
\ee
which correctly tends to (\ref{an1}) for $\tau \rightarrow 0$. In Fig. \ref{UCNAFig} we show the $\tau$ dependency of the UCNA solution for this system. The agreement between the direct stochastic simulations and the UCNA solution is excellent. It should be noted that the UCNA solution appears to
be virtually indistinguishable from the white noise solution for $\tau \ll \tau_x \approx 1/k = 10^3$ s$^{-1}$, while for increasing values of $\tau$, getting closer to $\tau_x$, the UCNA solution and the white noise solution diverge substantially from each other. This implies that as long as $\tau \ll \tau_x$, the white noise approximation works well even if the noise is characterized by a finite timescale, what we have named the 'virtually white' regime.   

We notice that despite the fact that the changes seen in probability distributions are limited to shifts of the position of the mode, and to its width, no change in the number of stationary states in the system occurs. This reproduces qualitatively the results \cite{Shahrezaei08}, where a similar shift was observed when the system is driven by log-normal degradation noise, and no noise-induced transitions were reported. We will discuss this further in Section 6. Similarly, the bimodality emerging in \cite{Holehouse20} for a self-activating gene is related to the pre-existent bistability of the system, with the noise changing the way the different (deterministic) attractor states are reached, through excursions in the bistable parameter regime. These conclusions are in agreement with the observation that extrinsic noise acting on parameters appearing linearly in the system is not expected to modify its dynamics substantially, and to cause transitions from unimodal to bimodal behaviours \cite{Dobrzynski14}.

\section{4. Non-Gaussian noises: Log-normal noise as a special case}
While the Central Limit Theorem would suggest the modelling of extrinsic fluctuations as a Gaussian process, this choice is questionable for fluctuations affecting a strictly positive parameter. The need of preserving the positivity of noisy parameters has brought along a number of approaches to model non-Gaussian noise sources. For instance, so-called bounded noises have been introduced recently and have attracted attention in abstract and more applied settings. A first example is offered by the so-called sine-Wiener noise, initially introduced in \cite{Bobryk05}, and applied for instance in \cite{dOnofrio10a,dOnofrio10b} in tumor biology. A second example is offered by the Tsallis-Borland noise, which with an adequate choice of parameters corresponds to a Tsallis statistics \cite{Borland98,Wio04}. Other types of bounded noises, currently being investigated, include the Cai-Lin noise \cite{Cai96}, and the Kessler-S{\o}rensen noise \cite{Kessler99}.

A particular mention goes to both Gamma-distributed noise and log-normal noise. Gamma-noise emerges as intrinsic noise in simple two-stage models of gene expression \cite{Friedman06,Shahrezaei08b}. Remarkably, Gamma distributions are robust to slow heterogeneities in parameter values \cite{Taniguchi10}, and therefore are a good fit to experimental data, integrating over both intrinsic and extrinsic noise sources, as shown in single cell experiments \cite{Cai06,Taniguchi10}. Similarly, log-normal noise is also a good candidate to represent fluctuations on parameter values phenomenologically \cite{Rosenfeld05,Shahrezaei08}. It is supported by both strong experimental evidence \cite{Bengtsson05} and theoretical arguments \cite{Ham20}, and it has been recently investigated to interpret universal features in bacteria and yeast \cite{Salman12,Brenner15}.

In this paper we will focus our analysis on log-normal noise. We define
\be
\lambda \rightarrow \lambda(t) = \bar{\lambda} e^{\eta(t)} e^{-D/2}, \label{deflogR}
\ee
where $\eta(t)$ is the standard Ornstein-Uhlenbeck noise, with the Langevin representation
\be
\frac{d \eta}{dt} = -\frac{\eta}{\tau} + \sqrt{\frac{D}{\tau}}\xi(t), \label{OU}
\ee
with $\xi(t)$ zero average Gaussian white noise. The definition (\ref{deflogR}) guarantees that $\langle \lambda \rangle = \bar{\lambda}$, since $\langle e^{\eta(t)} \rangle = e^{D/2}$. The fluctuations on $\lambda$ can then be described by
\be
\frac{d\lambda}{dt} = -\frac{\lambda}{\tau} \left[\ln \lambda - \ln \bar{\lambda} + \frac{D}{2} \right] + \sqrt{\frac{D}{\tau}} \lambda \xi(t) \label{langlognorm}
\ee
and are characterized by the stationary log-normal distribution:
\be
w(\lambda) = \frac{1}{\sqrt{2\pi D}} \frac{1}{\lambda} \exp \! \left[-\frac{1}{2 D}\left(\ln \lambda - \ln \bar{\lambda} +\frac{D}{2}\right)^2\right]. \label{pR}
\ee
It should be noted that the log-normal distribution (\ref{pR}) is always unimodal. In particular $w(0)=0$ for all parameter values.

\section{5. Nonlinear noise filtering and its dynamical derivation for slow noise}

Ochab-Marcinek et al. \cite{Ochab-Marcinek10} have recently introduced nonlinear noise filtering as a useful methodology to study the response of a gene under the effect of an external signal. In general terms, let us consider two variables, $x$ and $y$, such that $y = f(x)$. We can think of $x$ and $y$ respectively as input and output of a dynamical system, and the function $f$ as the response function. If the input variable $x$ is stochastic, with a probability distribution given by $p(x)$, this will induce a probability distribution $w(y)$ for the output variable $y$, thanks to the relation $y = f(x)$. The distribution for $y$ is readily computed by invoking conservation of probability:
\be
|p(x) dx| = |w(y)dy|
\ee
which then implies:
\be
w(y) = \left|\frac{dx}{dy}\right| p(f^{-1}(y)) \label{OM}
\ee
where $x = f^{-1}(y)$.  

Eq. (\ref{OM}) is at the basis of the noise filtering approach. If the relation $f$ between input and output is linear, the two proabability distributions are trivially linked, but in case of a nonlinear response function $f$, the transformation (\ref{OM}) introduces qualitative changes in their functional forms. For instance, depending on the input distribution, highly nontrivial effects can be expected for the family of Hill functions 
\be
y = f(x) = \frac{\alpha}{\beta + x},
\ee
where $\alpha$ and $\beta$ are shape parameters. This is the form adopted in \cite{Ochab-Marcinek10} to describe the response of a repressed gene to a regulating transcription factor. Similar considerations hold for signalling networks \cite{Dobrzynski14}, and for recently proposed models of bacterial persistence \cite{Rocco13a}.

In this section we will propose a dynamical derivation of the noise filtering approach when the correlation time of the extrinsic noise is much longer than any other timescale in the system. Extrinsic noise is indeed characterized by slow fluctuations, showing memory and correlation over time scales comparable to the entire cell cycle, or multiples of it, in both mammalian and bacterial cells \cite{Rosenfeld05,Kaufmann07,Sigal06,Taniguchi10}. We will also address the treatment of nonlinear noise, highly problematic in the white noise case \cite{Horsthemke84}, but possible and well-defined for finite and large correlation times. The dynamical derivation presented here recovers in the appropriate limit the fundamental result of \cite{Ochab-Marcinek10} on noise filtering as a purely geometric construction, and will allow us to identifies the limit of validity of the approach.

Let us then consider the stochastic system:
\be
&&\frac{dx}{dt} = \frac{1}{\tau_x}f(x,\lambda) \label{redP}\\
&&\frac{d\lambda}{dt} = \frac{1}{\tau}\mu(\lambda) + \sqrt{\frac{D}{\tau}} \nu(\lambda) \xi. \label{noise}
\ee
As before, the dynamics for $x$ are specified by the function $f(x,\lambda)$, and are characterized by the timescale $\tau_x$, while the control parameter $\lambda$ exhibits fluctuations described by (\ref{noise}). The variable $\xi=\xi(t)$ is again  a Gaussian, zero average, white noise, with correlator given by (\ref{corr}). The functions $\mu(\lambda)$ and $\nu(\lambda)$ are kept generic for the moment, but can be chosen so as to reproduce different types of fluctuations, characterized by different distributions, and different correlation times. For instance, they can match the Langevin equation (\ref{langlognorm}) to reproduce log-normal noise. 

Instead of studying the neighbourhood of white noise, for $\tau \rightarrow 0$,  in this section we are rather interested in developing a formalism to analyze the limit $\tau \gg \tau_x$. The rescaling by the factors $1/\tau$ in (\ref{noise}) is introduced so as to slow-down the fluctuations and allows us to carry out a systematic expansion in powers of $1/\sqrt{\tau}$, for $\tau \gg \tau_x$.

The Master Equation of the system specified by Eqs. (\ref{redP}) and (\ref{noise}) results in
\be
\frac{\partial w_t(x,\lambda)}{\partial t} = -\frac{\partial}{\partial x} \left[f(x,\lambda) w_t(x,\lambda)\right]
-\frac{1}{\tau}\frac{\partial}{\partial \lambda} \left[\left(\mu(\lambda) + D \nu(\lambda)\nu^\prime(\lambda)\right)w_t(x,\lambda)\right]
+ \frac{D}{\tau} \frac{\partial^2}{\partial \lambda^2} \left[\nu^2(\lambda) w_t(x,\lambda) \right], \label{ME}
\ee
where $w_t(x,\lambda)$ is the joint probability density for the $x$ and $\lambda$ variables. The term proportional to $\nu(\lambda)\nu^\prime(\lambda)$ is the so-called Stratonovich drift, corresponding to having assumed the Stratonovich interpretation for the multiplicative noise equation (\ref{noise}).

As it stands, Eq. (\ref{ME}) is difficult to solve, but it can be solved in stationary conditions when the stochastic fluctuations are slow. By following \cite{Arnold78,Reichl82}, we apply the so-called ``switching-curve approximation'', and expand the stationary solution $w_S(x,\lambda)$ of (\ref{ME}) in inverse powers of the correlation time $\tau$ of the $\lambda$ process:
\be
w_S(x,\lambda) = w_0(x,\lambda) + \frac{1}{\tau^{1/2}} w_1(x,\lambda)
+ \frac{1}{\tau} w_2(x,\lambda) + {\cal{O}}\left(\frac{1}{\tau^{3/2}}\right).
\ee
We then obtain to the zeroth order in $1/\tau$ 
\be
w_0(x,\lambda) = w(\lambda) \delta(x-u(\lambda)), \label{sol0}
\ee
where $u(\lambda)$ is defined so that $f(u(\lambda),\lambda)=0$. By assuming that $w_0(x,\lambda)$ is normalized, $w(\lambda)$ in (\ref{sol0}) can be identified by solving to order $\tau^{-1}$ the stationary Fokker-Planck Equation associated to (\ref{noise}) and supplemented with the Stratonovich interpretation:
\be
0 = D \frac{\partial^2}{\partial \lambda^2} \left[\nu^2(\lambda) w(\lambda) \right] 
- \frac{\partial}{\partial \lambda} \left[\left(\mu(\lambda) + D \nu(\lambda) \nu^\prime(\lambda)\right) w(\lambda)\right]. \label{wR}
\ee
The marginalized probability density $p(x)$ for the $x$ process can then be obtained as
\be
p(x) = \int w_0(x,\lambda) d\lambda = \int w(\lambda) \delta(x-u(\lambda)) d\lambda.
\ee
By using $\delta(\varphi(x)) = \delta(x-x_0)/|\varphi^\prime(x_0)|$, where $x_0$ is the root of $\varphi(x)$, $\varphi(x_0)=0$, we readily obtain:
\be
p(x) = w(u^{-1}(x)) \left| \frac{d u^{-1} (x)}{dx} \right|. \label{covs}
\ee
The probability density $w(\lambda)$ can be computed from (\ref{wR}), and we obtain 
\be
\hspace{-0.2cm }p(x) \! = \! \frac{{\cal N}}{\nu(u^{-1}(x))}
\!\left| \frac{d u^{-1} (x)}{dx} \right|\!
\exp \! \left\{\!\frac{1}{D} \!\! \int^{u^{-1}(x)} \!\! \frac{\mu(y)dy}{\nu^2(y)}\!\right\}, \label{sol}
\ee
where ${\cal N}$ is a normalization constant.

The mode(s) of the probability density $p(x)$ represent the natural extension of the stable steady state of the corresponding deterministic system, and can be computed explicitly by differentiating $p(x)$. This leads readily to
\be
\mu - D \nu \nu^\prime - D \nu^2 \left(\frac{d u^{-1}(x)}{dx}\right) u^{\prime \prime} = 0, \label{newStrat}
\ee
where $\mu=\mu(u^{-1}(x))$, $\nu=\nu(u^{-1}(x))$, and $u^{\prime \prime} = u^{\prime \prime}(u^{-1}(x))$.

Eq. (\ref{newStrat}) is an extension of the standard equation identifying the modes of a multiplicative stochastic process supplemented with the Stratonovich prescription, Eq. (\ref{modes}). The extra term $-D \nu^2 \left(d u^{-1}(x)/dx \right) u^{\prime \prime}$ in (\ref{newStrat}) accounts for dynamical modifications of the deterministic system which are induced by the slow fluctuations. This term is identically zero only in the case when the the relationship between the variable $\lambda$ and the variable $x$ is linear (so that $u^{\prime \prime} = 0$), while in all other cases nontrivial modifications of the dynamics can be expected, further and different from those arising in the white noise limit because of the Stratonovich drift. Again, whether these modifications will correspond to a shift of parameter values, a change of stability properties of the deterministic attractor states, or a change in their number (namely a purely noise-induced transition) will need to be assessed case by case, once the dynamics of the $x$ and $\lambda$ processes are specified.

As expected, Eq. (\ref{covs}) is equivalent Eq. (\ref{OM}). Our derivation highlights that the noise filtering approach is in fact valid in the case when no finite-time internal dynamics is present for the $x$ variable, namely in the case of infinite timescale separation $\tau/\tau_x \rightarrow 0$, and $x$ adapts istantaneously to the value dictated upon it by the dynamics of $\lambda$. This is the essence of the geometric construction presented in \cite{Ochab-Marcinek10}. 

It should be noted that the approach here presented has the advantage that it can be applied also in those cases when the driving distribution $w(\lambda)$ is not known. Eq. (\ref{newStrat}) allows us to identify the modes of the probability distribution $p(x)$ starting directly from the Langevin description of the process $\lambda$, Eq. (\ref{noise}), without requiring the analytical solution of its stationary probability density (\ref{sol}). 

\section{6. The repressed gene: Slow log-normal fluctuations on degradation}

Shahrezaei et al. \cite{Shahrezaei08} consider log-normal fluctuations on the degradation rate in a simple model of gene expression, and show that no noise-induced transition takes place in this case. Let us examine this case again by using our model system of the repressed gene.

For $\tau\gg\tau_x$ we can apply the results derived in the previous section and explore the effect of slow log-normal noise on the degradation rate $k$. By formally setting $\lambda \equiv k$, the funtions $\mu$ and $\nu$ entering Eq. (\ref{noise}) are thus identified as
\be
&&\mu(k) = k (\ln k - \ln \bar{k} + D/2)\\
&&\nu(k) = k
\ee 
with the function $u(k)$ resulting trivially in $u(k)=\beta/k$. Eq. (\ref{sol}) can be easily integrated, and results in:
\be
p(x) = \frac{1}{\sqrt{2 \pi D}}\frac{1}{x} \exp \! \left\{-\frac{1}{2 D} \left[\ln\left(\frac{\beta}{x}\right) - \ln \bar{k} +\frac{D}{2} \right]^2\right\}. \label{almostlognorm}
\ee

The resulting $p(x)$ is nontrivial in this case, as the original log-normal noise is transformed nonlinearly into the non-Gaussian probability density  (\ref{almostlognorm}). Despite this, the extrema are given by the equation
\be
x_M = \frac{\beta}{\bar{k}} e^{-D/2}, \label{single}
\ee
which is single-valued for all parameter values. 

\begin{figure}[t]
\begin{center}
  \hspace*{-0.5cm}\includegraphics[height=5.0 cm, angle=0]{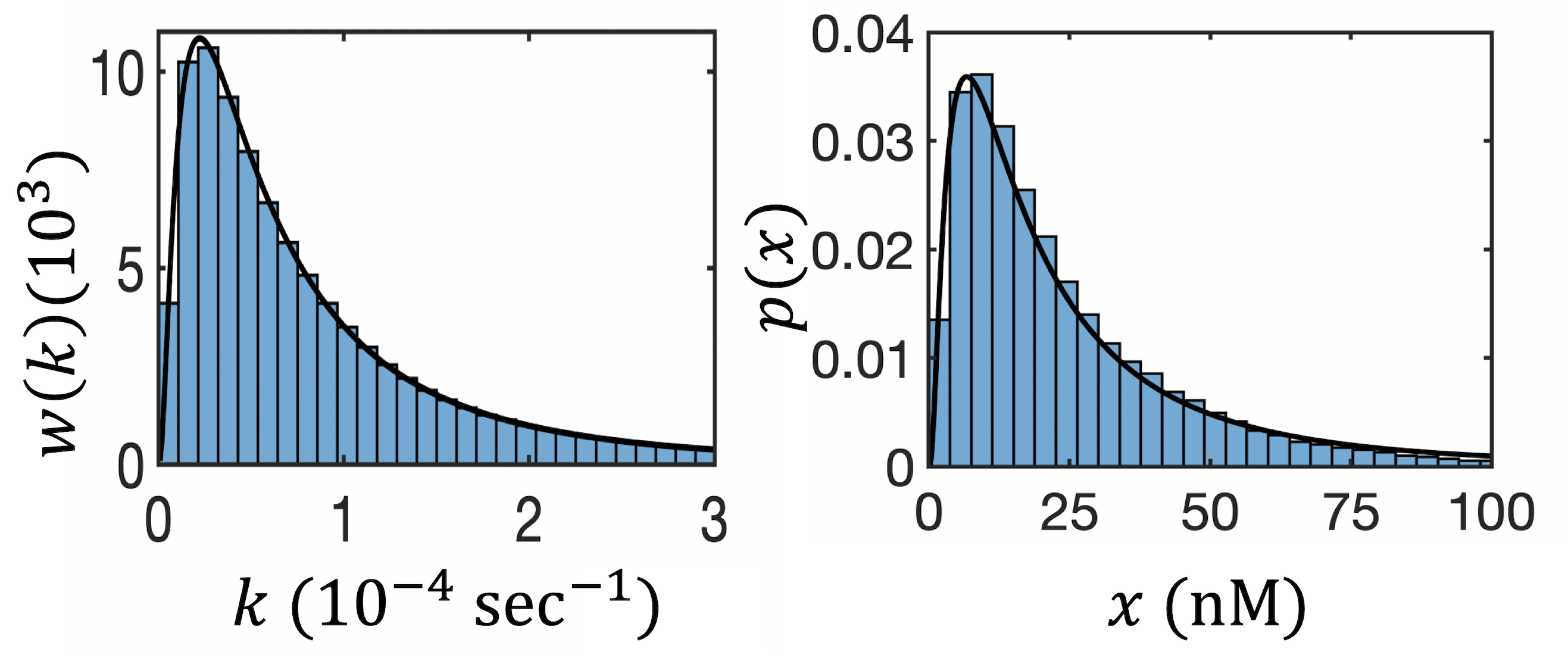} 
  \caption{Effect of log-normal noise on the degradation rate of the repressed gene. Left panel is the input log-normal dstribution, while the right paenl shows the $p(x)$. Parameter values are as follows: $g=0.01$ s$^{-1}$, $\rho=10$ nM$^{-1}$, $\bar{k}=10^{-4}$ s$^{-1}$, $\tau=10^5$ s, $D=1$. The mode is located at $x_M=6.67$ (nM).} \label{kLogNormFig}
\end{center}
\end{figure}

Simulations confirm nicely these analytical predictions, as shown in Fig. \ref{kLogNormFig}. In particular the single solution given by (\ref{single}) implies that no transition is possible in this system, despite our choice of log-normal noise to model fluctuations on $k$. It should be noted that the same result reported in \cite{Shahrezaei08} is derived by adopting the UCNA approach, and therefore valid for the correlation time of the noise being either very small or very large.

\section{7. The repressed gene: Fluctuations on the repressor}

Let us now come to the core of the nonlinear noise filtering approach \cite{Ochab-Marcinek10}. By using again our model system, the repressed gene, in this section we want to highlight with concrete examples the two main factors of the nonlinear noise filtering approach responsible for the occurrence of noise-induced transitions: The nonlinear dependency of the gene dynamics on fluctuating parameters, and the input distribution of the fluctuations.

\begin{figure}[t]
\begin{center}
  \includegraphics[height=5.0 cm, angle=0]{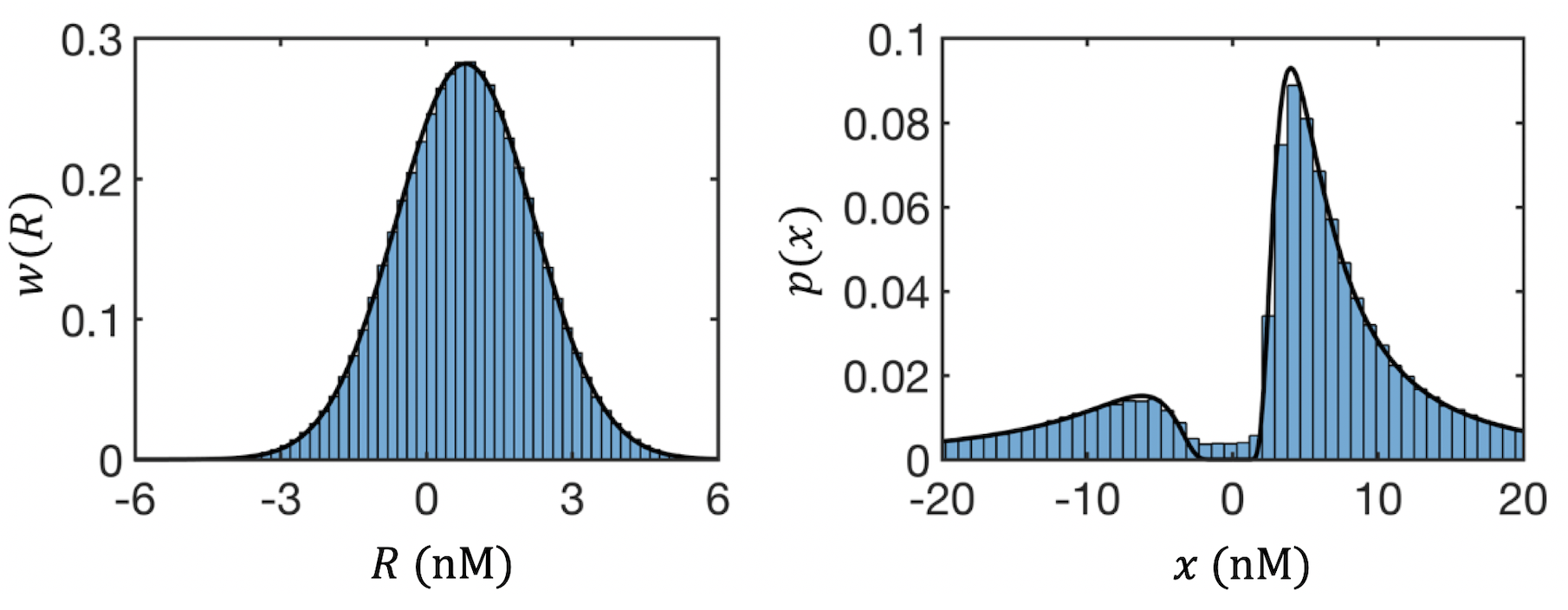} 
  \caption{Ornstein-Uhlenbeck Gaussian distribution for $R$ and corresponding probability distribution for $x$ as from Eqs. (\ref{pxOU}) for $D=2$. Other parameters: $g=0.01$ s$^{-1}$, $\rho=10$ nM$^{-1}$, $k=10^{-4}$ s$^{-1}$, $\bar{R}=0.8$ nM, $\tau=10^5$ s.} \label{OUnot}
\end{center}
\end{figure}

\begin{figure}[h!]
\begin{center}
  \hspace*{-0.5cm}\includegraphics[height=5.0 cm, angle=0]{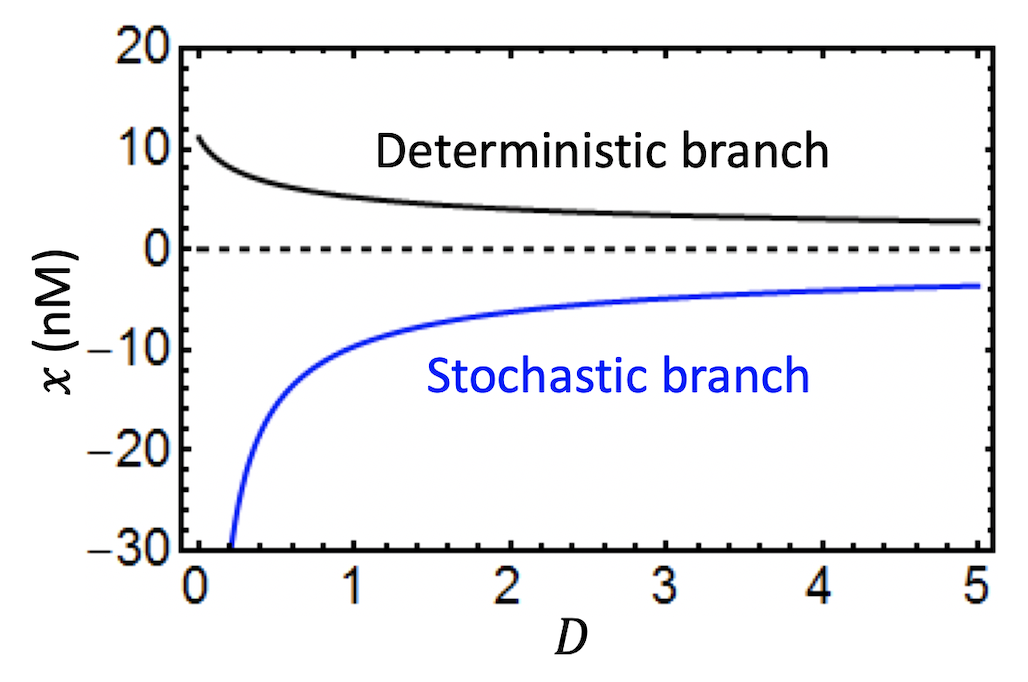} 
  \caption{Corresponding bifurcation diagram. The deterministic branch emanates from the deterministic solution at $D=0$. The stochastic branch only exists for $D>0$. Parameter values as in Fig. \ref{OUnot}.} \label{OUnot2}
\end{center}
\end{figure}

It is worth noticing that already in this simple system, different choices can be made for the identification of the nonlinear relevant parameter of interest, which correspond to different biological interpretations. For instance, Ochab-Marcinek et al.\cite{Ochab-Marcinek10} focus on fluctuations in the relevant transcription factor, in our case the repressor $R$, while Birtwistle et al \cite{Birtwistle12} and Dobrzy{\`n}ski et al. \cite{Dobrzynski14} focus instead on fluctuations on activation thresholds. Both choices are legitimate, but correspond clearly to different biochemical processes. An important question here is how to justify the assumed distribution of fluctuations on the relevant parameters. We will discuss this specific point in our Concluding Remarks, Section 8. 

By following \cite{Ochab-Marcinek10} we make the choice of considering slow fluctuations acting on $R$. First we will consider Gaussian noise, and then we will address log-normal noise.

In the case of Gaussian fluctuations acting on the parameter $R$, we assume $R \rightarrow \bar{R} + \eta(t)$, where $\eta(t)$ is the standard Ornstein-Uhlenbeck process, whose dynamics is given by Eq. (\ref{OU}). Eq. (\ref{sol}) can be readily integrated once the function $u(R)$ is defined as
\be
\frac{g}{1+\rho R} - k u(R) = 0 \quad \Rightarrow \quad u^{-1}(x) = \frac{g}{k \rho x} - \frac{1}{\rho}.
\ee
By direct integration of Eq. (\ref{sol}), we readily obtain
\be
p(x) \!= \! \frac{1}{\sqrt{2 \pi D}} \! \left(\frac{g}{k \rho x^2} \right) \! \exp \! \left\{-\frac{1}{2 D} \left( \frac{g}{k \rho x}\! -\! \frac{1}{\rho}\! -\! \bar{R} \right)^2\right\}, \label{pxOU}
\ee
again showing a nonlinear tranformation from the original Gaussian noise. 

Simulations agree again very well with the analytical predictions, as it it is shown in Fig. \ref{OUnot}. The timescale of the fluctuations $\tau$ is crucial for the matching of our theoretical predictions with the numerical results. Given $k=10^{-4}$ s$^{-1}$, we set $\tau = 10^5$ s, to capture the dynamics of slow fluctuations of $R$. 

However, only one positive mode is present, which extends the deterministic solution. In fact, for $x \neq 0$, Eq. (\ref{newStrat}) becomes
\be
2D k^2 \rho^2 x^2 + g(1+\rho \bar{R})kx - g^2 = 0 \label{zerosgauss}
\ee
which for $D \neq 0$ accounts for the two real solutions:
\be
x_{1,2} = -\frac{g}{4 D k \rho^2}(1+\bar{R} \rho) \pm \sqrt{\left[\frac{g}{4 D k \rho^2}(1+\bar{R} \rho)\right]^2 + \frac{g^2}{2 D k^2 \rho^2}}.
\ee

The solution of (\ref{zerosgauss}) for $D=0$ coincides trivially with the solution of the deterministic case, with $x_1$ being the stochastic continuation of it (Fig. \ref{OUnot2}). The $x_2$ branch emerges instead for $D>0$, signalling that no transition is taking place in the system at finite $D$ (the only critical point being the trivial one, $D=D_c=0$). Despite the appearance of a second mode, this is to be biologically discarded because at negative $x$ values.

\begin{figure}[t]
\begin{center}
  \includegraphics[height=13.5 cm, angle=0]{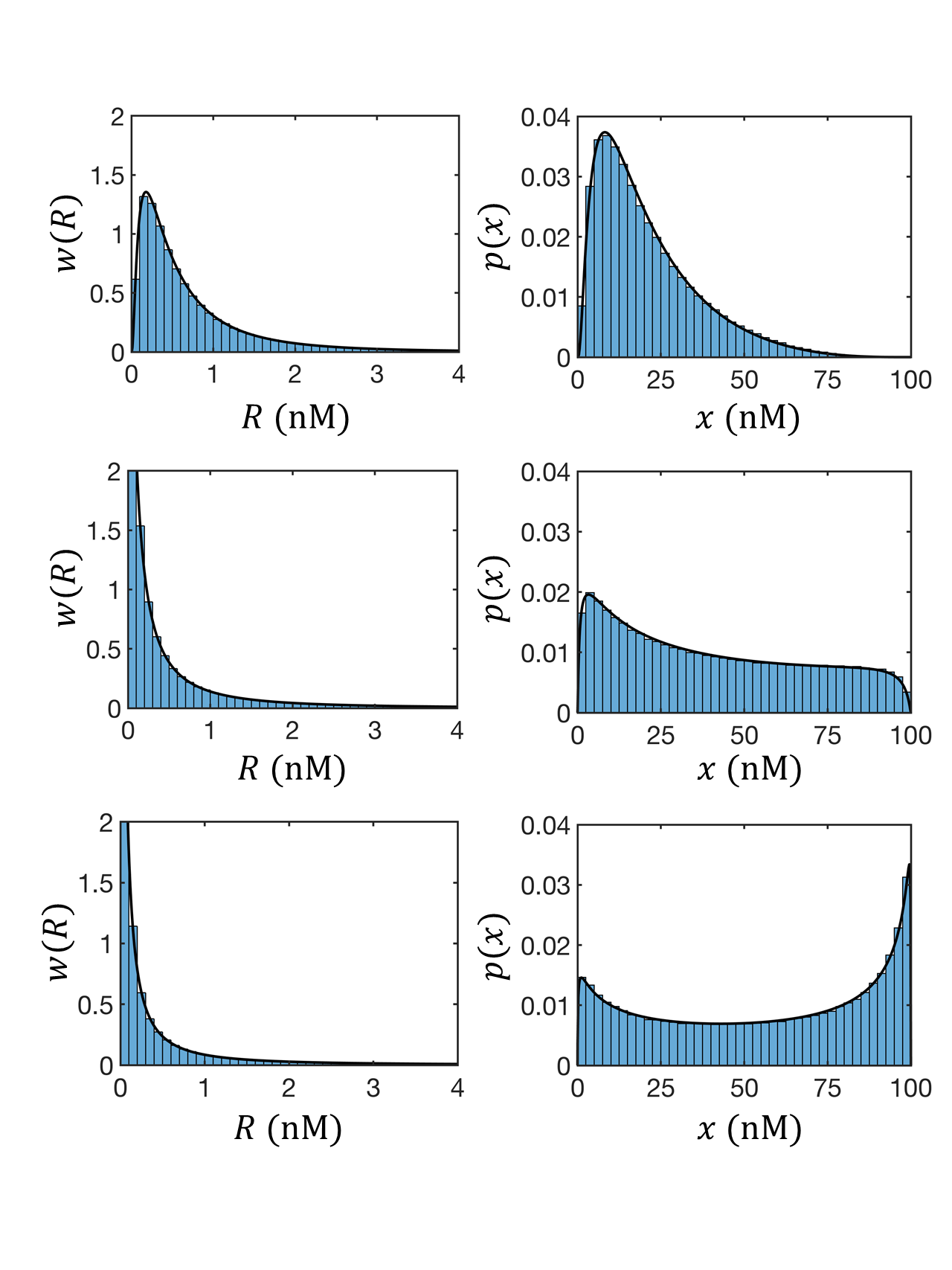} 
\vspace*{-1.5cm}
  \caption{Log-normal distributions for $R$ and corresponding probability distributions for $x$ as from Eqs. (\ref{pR}) and (\ref{px}) respectively for different values of noise intensity $D$. From top to bottom, $D=1$, $D=3$, and $D=5$. For $D=5$, modes are located at $x_1=1.14$ and $x_2=99.5$ (nM). Parameter values as in Fig. \ref{OUnot}.} \label{trans}
\end{center}
\end{figure}

Let us consider now log-normal noise. By direct integration of Eq. (\ref{sol}), we readily obtain
\be
&&\hspace{-1cm}p(x) = \frac{g}{\sqrt{2\pi D}} \; \frac{1}{x(g - k x)} \nonumber \\
&&\exp\left \{-\frac{1}{2 D}\left[ \ln\left(\frac{g}{k \rho x} -\frac{1}{\rho} \right) - \ln \bar{R} +\frac{D}{2} \right]^2\right\}\label{px}
\ee
and, for $x\neq 0$ and $x \neq g/k$, from Eq. (\ref{newStrat})
\be
g \ln \left(\frac{g}{k \rho x} -\frac{1}{\rho} \right) - g \ln \bar{R} - \frac{gD}{2} + 2Dkx = 0. \label{zeros}
\ee

\begin{figure}[t] 
\begin{center}
  \includegraphics[height=5.0 cm, angle=0]{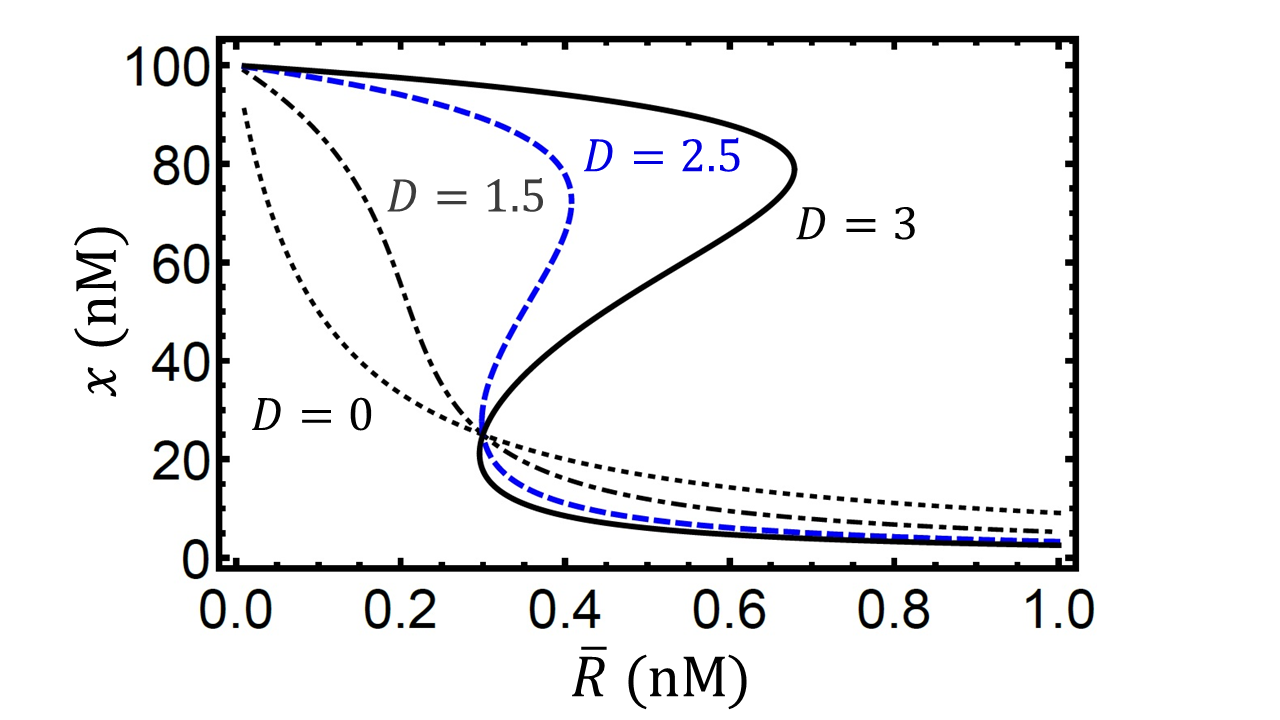} 
\caption{Stationary-state response curves for different values of the noise intensity, as determined by Eq. (\ref{zeros}). The case $D=0$ corresponds to the deterministic system. Parameter values as in Fig. \ref{OUnot}.} \label{bifdia}
\end{center}
\end{figure}

Simulations show an excellent agreement with the theoretical predictions, as shown in Fig. \ref{trans}. Further to the excellent matching of the full distribution for $R$, the predictions for the modes as given by Eq. (\ref{zeros}) is also verified. In Fig. \ref{bifdia} we show the extrema of the stationary probability (\ref{px}) as function of $\bar{R}$ for different noise intensities, obtained by the numerical solution of Eq. (\ref{zeros}) with parameter values as in Fig. \ref{trans}. The curve associated to $D=0$ (deterministic case) is single valued for all values of $\bar{R}$, with further increase of $D$ provoking the appearance of a second stable mode for intermediate $\bar{R}$ values.  

These results confirm that the emergence of noise-induced transitions in simple regulatory dynamics depends strongly on the type of noise and on the nonlinearity involved. From a biological perspective it is then essential to assess carefully noise distributions and network reconstructions to be able to make clear predictions. Certainly non-Gaussian noises, in the case explored here log-normal, long fluctuations correlations, and consideration of nonlinear parameters, seem to be essential ingredients for the emergence of noise-induced transitions.   

\section{8. Concluding remarks}

In this paper we have summarized how the concept of noise-induced transitions has evolved over the last few decades, and has now become an active area of research in gene systems and molecular cellular processes in general. 

With the help of a simple gene expression system, the repressed gene, we have reviewed different approaches to model extrinsic fluctuations, ranging from assuming Gaussian white and coloured extrinsic noise, to the adoption of bounded noises \cite{deFranciscis14} and nonlinear noise filtering methodologies \cite{Ochab-Marcinek10}. The repressed gene model offers scope for interesting analysis as it is defined in terms of parameters that appear in the dynamical equations both linearly and nonlinearly, a fact that is now known to contribute differently to transitions to bimodality \cite{Dobrzynski14}. In this paper we have reviewed the behaviour of this systems with respect to its linear dependency on the degradation rate $k$, and the nonlinear dependency of the repressor concentration, $R$.

When the noise appears linearly in the repressed gene, namely on the degradation rate, standard approaches can be adopted to carry out analytical predictions in the case of Gaussian white noise. These standard tools, extended to coloured noise by the powerful UCNA approach \cite{Jung87}, and adopted in a variety of different gene regulatory systems \cite{Shahrezaei08,Holehouse20}, do not predict noise-induced transitions. The same result holds true for log-normal noise characterized by long correlation time \cite{Shahrezaei08} .

However, when fluctuations appear in a nonlinear fashion, noise-induced transitions become possible. From a mathematical perspective, treating nonlinear noise dynamically is usually a hard problem, but recent progress has made it possible by adopting the noise filtering approach introduced in \cite{Ochab-Marcinek10,Ochab-Marcinek17}. We have here reviewed this approach, and proposed a dynamical derivation of it in the limit of large correlation time of the noise. The key equation describing the approach (Eq. (\ref{OM}) or Eq. (\ref{covs})) highlights how nontrivial behaviours, including transitions to bimodality, depend on the combined effect of two dynamical features. The first one is represented by the derivative term in Eqs. (\ref{OM}) and (\ref{covs}), which represents the response function of the system, whose nonlinearity is responsible for the distortion of the input probability distribution. For linear filtering, this derivative is just a numerical factor, and hence ininfluent in modifying dynamics between input and output. The second important ingredient is of course the nature of the input distribution itself. We show here with a concrete example based on the repressed gene that if the input distribution is Gaussian no biologically meaningful bimodality is obtained (a second mode indeed appears, but it is at negative concentrations), while if it is log-normal a physical transition occurs, with two positive mode emerging. The same analysis holds for Gamma distributed noise, as presented in \cite{Ochab-Marcinek10,Kim12,Birtwistle12}.

Whether the Gaussian noise approximation for extrinsic noise (either white or coloured) is biologically feasible is a delicate question. While it is indeed true that the Gaussian distribution produces negative values for fluctuating parameters, it is unquestionable that it has got its own merits, which also rely on the development of poweful and thoroughly tested analytical results. In fact in recent papers, such as \cite{Shahrezaei08} and \cite{Holehouse20} the Gaussian approximation has been carried out, leading to clear and reliable predictions on the behaviour of the system. It is fair to say that according to the the system itself, and the type of noise one wants to model, the Gaussian approximation may still be an attractive and valid alternative. Current experimental results however, support the relevance of bounded noises to fit models to data. For instance, the already mentioned studies \cite{Birtwistle12,Kim12} and \cite{Dobrzynski14} show an excellent matching of models to experiments by adopting Gamma noise and the nonlinear filtering approach, which allows for the identification of the related noise-induced transitions.

A number of challenging problems remains open. One important one concerns the interplay between intrinsic and extrinsic noise. Intrinsic and extrinsic noises are invitably present in the system, and ideally should be both considered in the models proposed. Several attempt have been made to tackle this issue, with promising results. For instance, in \cite{Lei09} the chemical Lengevin equation with the It\^{o} prescription has been adopted to model intrinsic fluctuations, while extrinsic noise is treated with the Stratonovich prescription. A similar approach has been put forward in \cite{Holehouse20} for the self-regulating gene, and in \cite{Maleki17} in the context of reducing model dimensionality of feedback loops, by adopting the so-called loop opening approach defined in \cite{Angeli04,Hsu16}. It is worth noticing that in all these cases the SDEs obtained are 'doubly' multiplicative, in that both noise sources provoke the apperance of multiplicative noise terms. Since intrinsic fluctuations (white) and dynamical variables are uncorrelated, for these the It\^{o} prescription follows, while the fact that extrinsic fluctuations (coloured) develop time correlations with the dynamical variables is reflected in the adoption of the Stratonovich prescription. Because of these rich dynamics, noise-induced transitions are likely, and indeed emerge in both studies \cite{Lei09,Holehouse20}.

A second open problem concerns the origin of the different input noise distributions chosen to model experimental data and noise-induced transitions. A range of different assumptions has been made. For instance, fully bounded distributions have been postulated to describe tumor biology \cite{dOnofrio10a,dOnofrio10b}, while semibounded noises (Gamma and log-normal distributed) have been postulated in the studies \cite{Ochab-Marcinek10,Birtwistle12,Dobrzynski14} either on activation thresholds, or on regulating transcription factors. In the case of a truly external parameter, like an activation threshold, this choice is made on a phenomenological basis, and it may be hard to justify it in terms of molecular processes. However, fluctuations affecting a transcription factor deserve more attention. These may in fact have an intrinsic origin, as they come about through the dynamical processes involved in gene expression, as shown in \cite{Friedman06,Shahrezaei08b}, where it is predicted that fluctuations in gene expression follow a Gamma distribution, a result confirmed experimentally in \cite{Cai06}. In the approach adopted in \cite{Ochab-Marcinek10}, and followed by ourselves in this review, it is crucial to treat the transcription factor as an external parameter in the dynamics of the target gene, so that adopting a stationary Gamma (or log-normal) distribution to describe its fluctuations is fully justified. However, a full dynamical treatment would require modelling in terms of a 3-dimensional system, accounting for transcription factor production, target gene transcription factor-mediated regulation, and noise dynamics. To reduce this 3-dimensional system to a 2-dimensional system (target gene dynamics with the transcription factor appearing as an external parameter, plus noise), as \cite{Ochab-Marcinek10} and we do, may be a delicate step. While achieving this reduction is mathematically manageable for deterministic systems, how to realize it consistently for a stochastic system is more challenging. This requires eliminating fast variables by stochastic adiabatic elimination \cite{Kupferman04,Pavliotis05} or projection techniques, a possibility currently under investigation \cite{Thomas12,Bravi20}. In this sense, intrinsic noise could in fact behave as extrinsic and be ultimately responsible itself for nontrivial behaviours \cite{Perez-Carrasco16}, with the boundary between intrinsic and extrinsic dynamics becoming fuzzy and dependent on the definition of networks and subnetworks.

A third problem concerns the nature and the implications of the observed bimodality. Even though the population can be distributed in two or more subpopulations in terms of gene profiles, the internal dynamics of switching between them will be relevant for drawing biological conclusions. The same bimodal distribution can be realized in quasi-static terms, with limited transitions between states, or very fast transitions between them, or anything in between. The timescale of the 'hopping' process will be relevant to assign observable phenotypes to those gene configurations corresponding to specific states. Only slow hopping corresponds to what is commonly referred to as a clear switching behaviour, or, mathematically speaking to (possibly weak) ergodicity breaking \cite{Rocco13a}. Hysteresis, as a signature of true bimodal dynamics, can be seen in the same way. P{\'a}jaro et al. \cite{Pajaro19} point out that indeed hysteresis may only have a transient nature under the presence of intrinsic noise, as eventually the system converges to equilibrium, and restores ergodic behaviour. This phenomenon is particularly relevant for slow fluctuations, when the rate of convergence to equilibrium is itself slow \cite{Brock09}, and needs to be studied case by case.

All these results indicate that particular care should be taken when reconstructing gene regulatory networks under the evidence of bimodal distributions of gene expression levels. The usual implication that these require a wiring diagram including feedback loops may be false, as the same bimodality may be produced by nonlinear extrinsic noise. It is intriguing to hypothesize that management of noise could provide an alternative way evolution adopts to realize bimodal physiology of cellular systems \cite{Sanchez13,Tsimring14}. Certainly the mechanisms for noise-induced transitions here reviewed are promising and worthy of further analysis and experimental validation. These will contribute to our  understanding of the fundamental relevance of noise in biological systems. 
  
\section*{Acknowledgements}

This work was supported by the UK Biotechnology and Biological Sciences Research Council (BBSRC) under grant BB/L007789/1. AR wishes to thank the Editors of this Special Issue for the invitation to submit this paper. 

\section*{Conflict of interest}

The authors declare no conflict of interest in this paper.

\end{document}